\documentstyle[15lomcon,cite,epsfig]{article}
\graphicspath{{ps/}}

\begin{document}

\title{STATUS AND PROSPECTS OF SUPERKEKB COLLIDER AND BELLE II EXPERIMENT}

\author{Tagir Aushev \footnote{e-mail: aushev@itep.ru}}

\address{Institute for Theoretical and Experimental Physics, 117218 Moscow, Russia}

\maketitle
\ \\
\abstracts{High precision measurements in the quark flavor sector are
  essential for searching for new physics beyond the Standard model.
  SuperKEKB collider and Belle II detector are designed to perform
  such measurements.  The status and prospects of the SuperKEKB and
  Belle II are presented in this article.}

\section{Introduction}

Since the end of the last century, two asymmetric-energy $e^+e^-$ $B$
factories, the KEKB~\cite{KEKB} collider for the Belle~\cite{Belle}
experiment at KEK and the PEPII collider for the BaBar experiment at
SLAC, have been achieving a tremendous success that lead to the
confirmation of the Standard Model (SM) in the quark flavor sector.
The main goal of the experiments was to measure the large
mixing-induced $CP$ violation in the $B^0$ system predicted by the
theory of Kobayashi and Maskawa~\cite{KM}.  The experimental data
indicated that the Kobayashi-Maskawa mechanism is indeed the dominant
source of the observed $CP$ violation in Nature. Following the
experimental confirmation, M.~Kobayashi and T.~Maskawa were awarded
the 2008 Nobel Prize for physics.

In addition to the observation of the $CP$ violation in $B$ meson
system, a numerous of other important measurements and observations
have been done by both experiments, such as measurements of all angles
of the unitarity triangle; direct $CP$ asymmetry
in $B^0\to\pi^+\pi^-$ and $K^+\pi^-$; the first observation of rare
$B$ decays, such as $B\to K^{(*)}\ell\ell$, $b\to s\gamma$ and
$\tau\nu$; observation of the new type of particle, such as $X(3872)$;
observation of the $D^0$ mixing, etc.

Most of the results are in a good agreement with the expectation from
the SM, however some measurements show tension with the SM prediction.
A significantly larger statistics is necessary to investigate whether
these are first hints for effects of a new physics.  To solve this
task a next generation experiment, Belle II, operating at the high
luminosity collider, SuperKEKB, is designed.  In this article, the
status and prospects of the SuperKEKB collider and Belle II detector
are presented.

\section{The Belle experiment}

The Belle detector was operating on the KEKB asymmetric-energy
$e^+e^-$ collider.  From 1999 to 2010, KEKB delivered an integrated
luminosity of about $1040{\rm~fb}^{-1}$.  Most of the data were taken
at the center-of-mass energy of the $\Upsilon(4S)$ resonance and
contains about 772 million $B$ meson pair events.  The achieved peak
luminosity is $2.1\times10^{34}{\rm~cm^{-2}s^{-1}}$.  On June 30,
2010, the Belle experiment was stopped with the ceremonial dump of the
last KEKB beam.  Currently, the Belle detector is rolled out from the
beam interaction point and partially disassembled.

\section{Hints for a new physics}

\begin{figure}[htb]
\includegraphics[height=5.5cm]{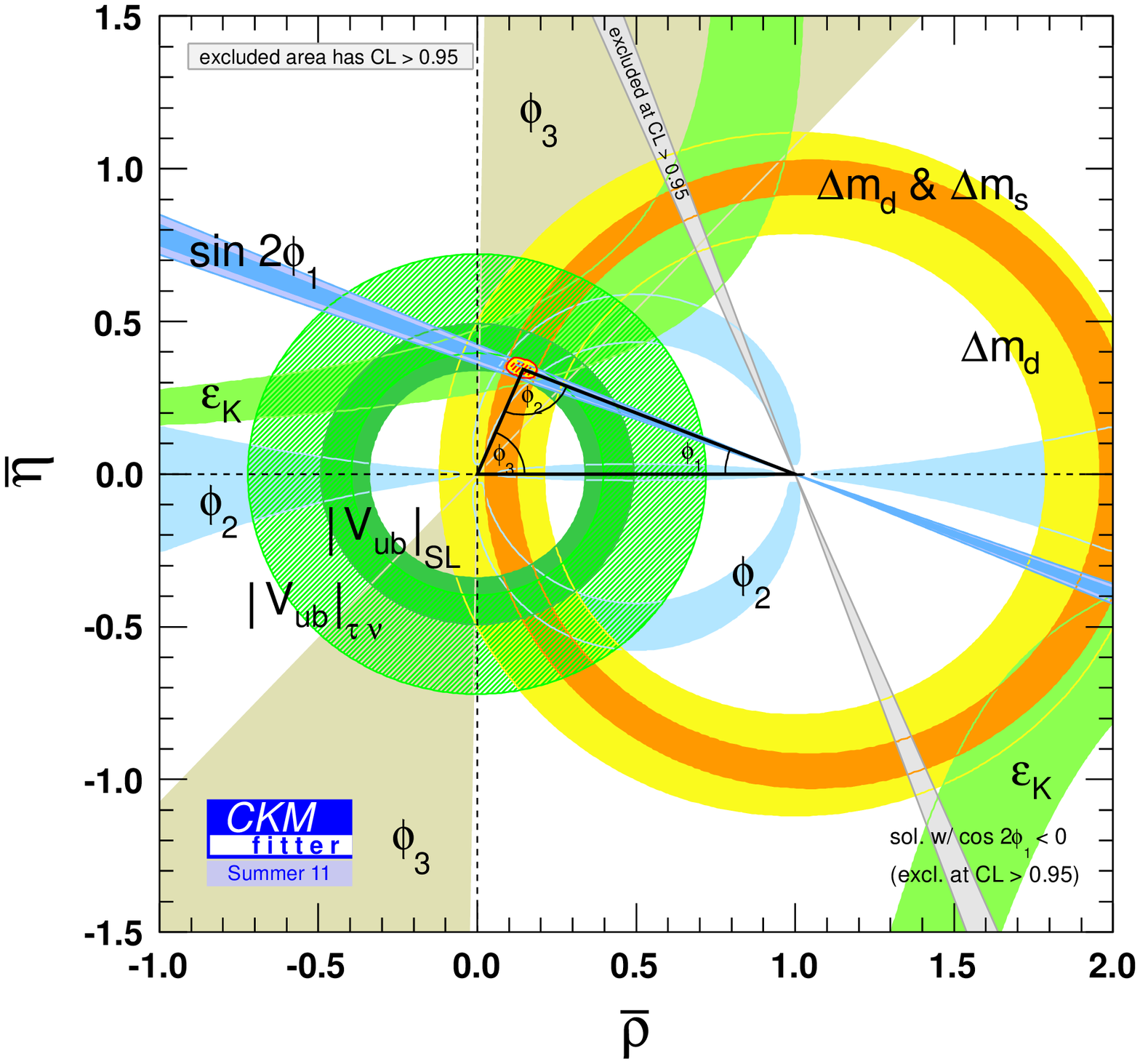}
\includegraphics[height=5.5cm]{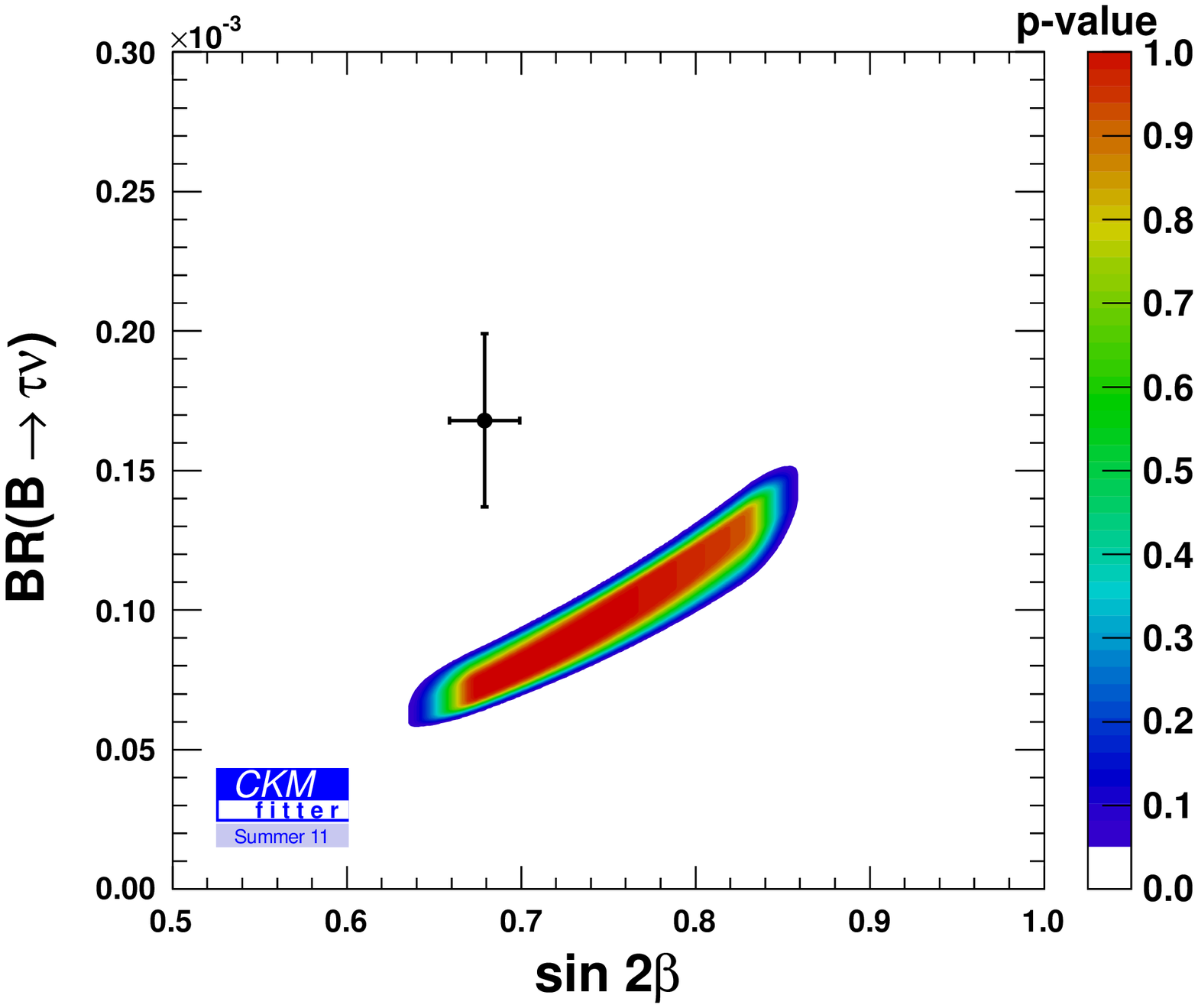}
\caption{Unitarity triangle status from the CKMfitter group (left) and
  the tension between the measurements of ${\cal B}(B\to\tau\nu)$ and
  $\sin2\phi_1$ with predictions (right).}
\label{triangle}
\end{figure}

Due to the unitarity of the Cabibbo-Kobayashi-Maskawa (CKM) 
matrix and complexity of its elements,
one can build a unitarity triangle on the complex plane from the CKM
matrix elements.  All sides and angles of this triangle can be
measured independently, and the consistency of the obtained results is
an important check of the SM and a search for a new physics.
Currently, most of the CKM parameters are well measured and a room for
a new physics is rather small (Fig.~\ref{triangle} left).

However, there are still some small discrepancies: the indirect
determination of the angle $\phi_1$ is exhibiting a $2.7\sigma$
deviation from the current world average 
of direct measurements of $\sin2\phi_1$~\cite{CKMfit}.
Equivalently, the $B^\pm\to\tau^\pm\nu_\tau$ branching fraction and
the resultant $|V_{ub}|$ show a deviation of $2.8\sigma$ from the one
predicted by the global fit~\cite{CKMfit}, where the $\sin2\phi_1$
value gives the most stringent constraint on the indirect
measurement (Fig.~\ref{triangle} right).

Another place where a new physics can reveal itself is the decay of $b\to
s\bar ss$.  In the SM a time-dependent $CP$ violation in the decay
$B^0\to\phi K^0_S$ is expected, similarly to $B^0\to J/\psi K^0_S$, to
be $\sin2\phi_1$.  However, the existence of new particles in the
penguin loop in the decay $B^0\to\phi K^0_S$ can deviate the observed
value from the expected one~\cite{bib_b2s_theory}.  The current
measurement gives $\Delta S\equiv\sin2\phi_1^{B\to\phi
  K^0_S}-\sin2\phi_1^{B\to J/\psi K^0_S}=0.22\pm0.17$.  The goal of
the next experiment is to reduce the error of this measurement by the
factor of 10.

A direct $CP$ violation was measured in $B\to K\pi$ system.  Since
both tree and penguin processes contribute to $B^0\to K^+\pi^-$ and
$B^+\to K^+\pi^0$ decays, sizable ${\cal A}_{CP}$ is expected.
Moreover, ${\cal A}_{CP}(B^0\to K^+\pi^-)$ and ${\cal A}_{CP}(B^+\to
K^+\pi^0)$ are expected to have approximately same magnitude and
sign~\cite{kpi_theory}.  Oppositely, both $B^+\to K^0\pi^+$ and
$B^0\to K^0\pi^0$ are almost pure penguin processes, hence no sizable
asymmetries are expected in the SM.

Consistent with no asymmetry results have been obtained experimentally
for the decays $B^+\to K^0\pi^+$ and $B^0\to K^0\pi^0$ as it was
expected from the theory.  However, the measured asymmetries ${\cal
  A}_{CP}(B^0\to K^+\pi^-)$ and ${\cal A}_{CP}(B^+\to K^+\pi^0)$ have
different magnitudes and signs, and their difference is $\Delta{\cal
  A}_{CP}={\cal A}_{CP}(B^0\to K^+\pi^-)- {\cal A}_{CP}(B^+\to
K^+\pi^0)=-0.147\pm0.28$, which has been established with a
significance of $5.3\sigma$.

\begin{figure}[htb]
\centering
\includegraphics[height=5cm]{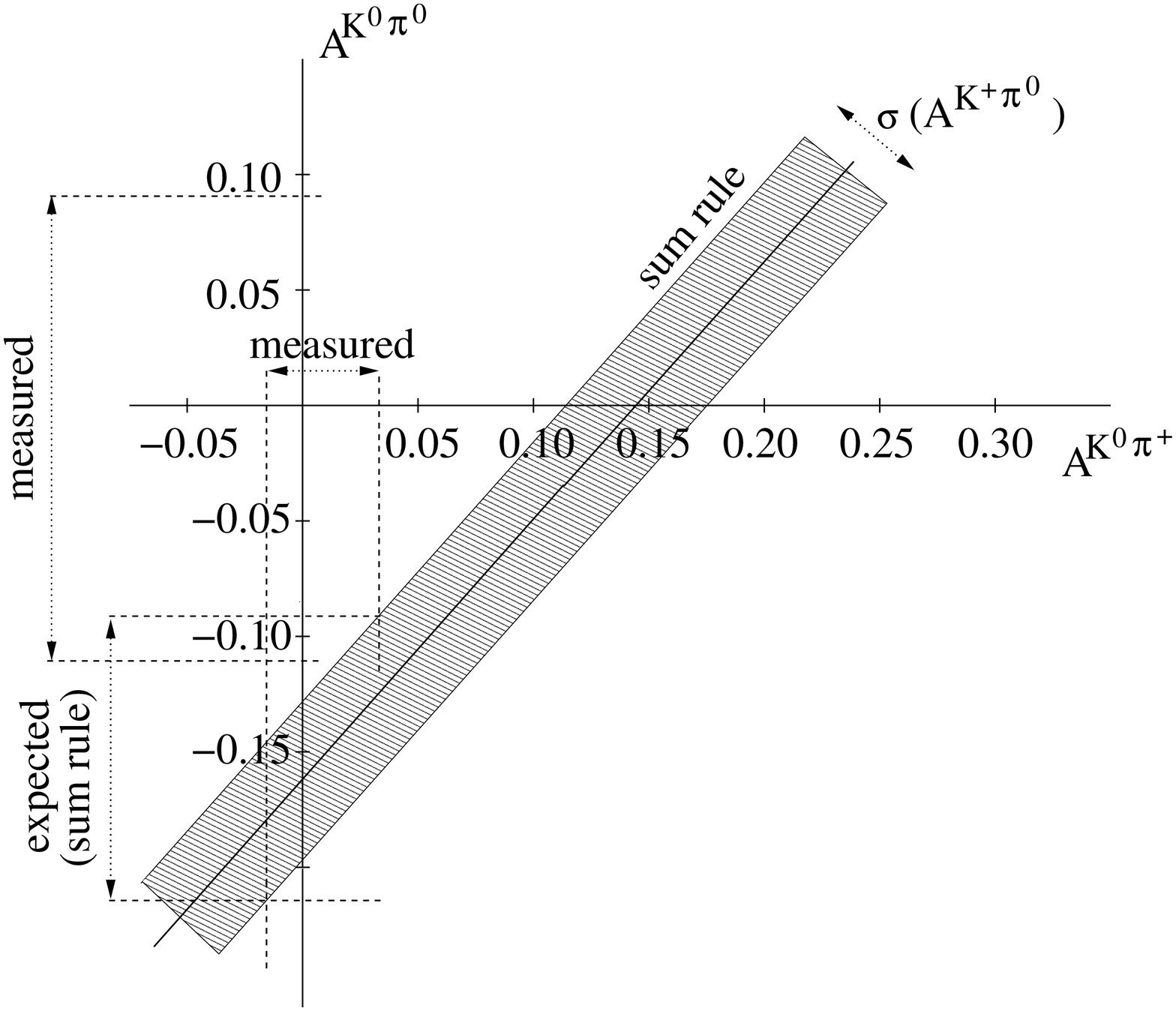}
\includegraphics[height=5cm]{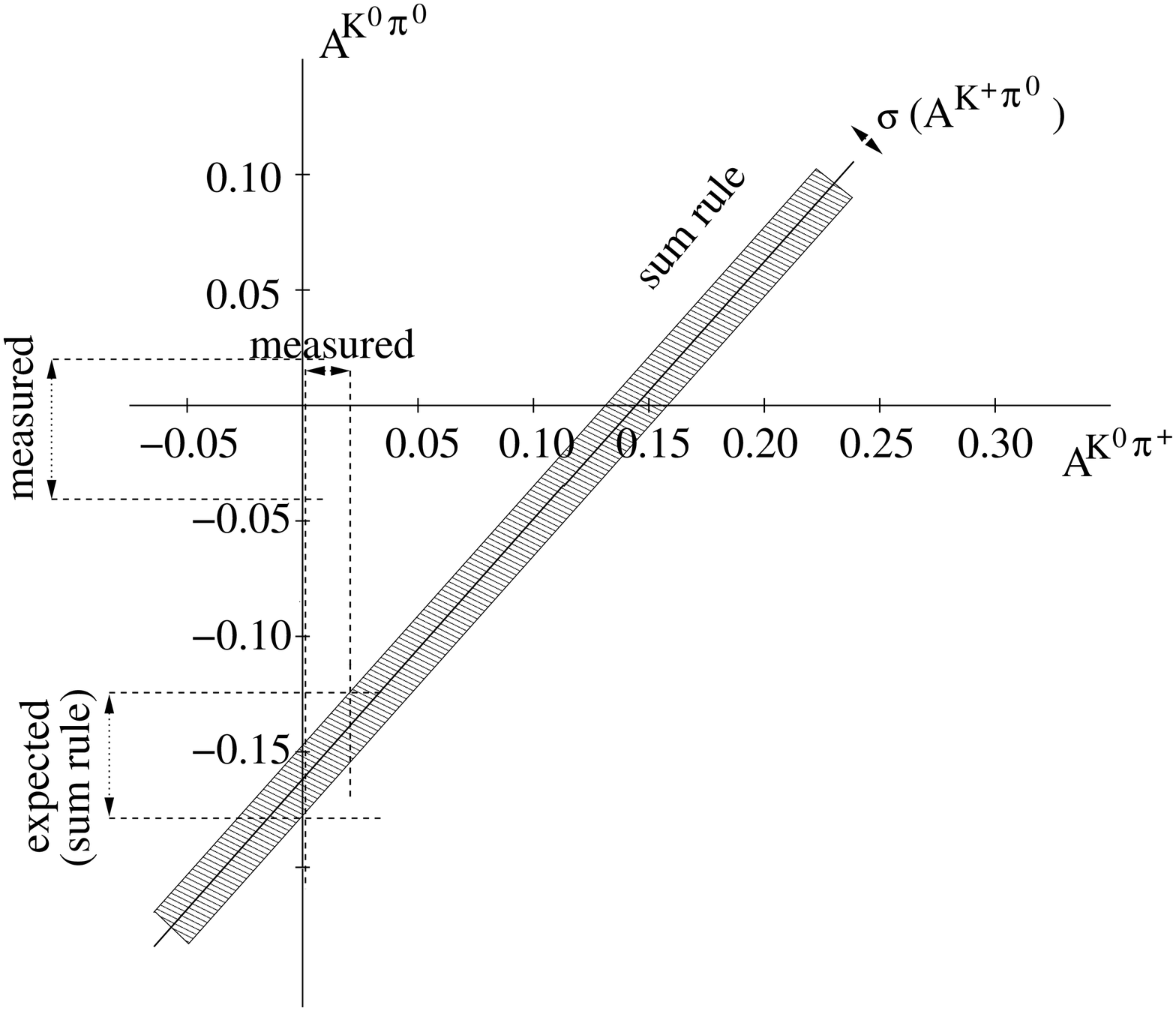}
\caption{Illustration of the sum rule for the current experimental
  values (left) and the projection for SuperKEKB assuming the same
  central values (right).}
\label{kpi_sum}
\end{figure}

There are several theoretical models, which explain the sizable
$\Delta{\cal A}_{CP}$ effect by the colour-suppressed tree and penguin
processes.  To exclude these effects and examine for a new physics,
the isospin sum rule among four ${\cal A}_{CP}$ values can be
applied~\cite{kpi_sum}:
\[
{\cal A}_{CP}^{K^+\pi^-}+{\cal A}_{CP}^{K^0\pi^+}
\frac{{\cal B}(B^+\to K^0\pi^+)\tau_{B^0}}{{\cal B}(B^0\to K^+\pi^-)\tau_{B^+}}=
\]
\[
{\cal A}_{CP}^{K^+\pi^0}
\frac{2{\cal B}(B^+\to K^+\pi^0)\tau_{B^0}}{{\cal B}(B^0\to K^+\pi^-)\tau_{B^+}}+
{\cal A}_{CP}^{K^0\pi^0}
\frac{2{\cal B}(B^0\to K^0\pi^0)}{{\cal B}(B^0\to K^+\pi^-)},
\]
where $\tau_{B^0}$ ($\tau_{B^+}$) is a $B^0$ ($B^+$) meson lifetime.
This relation is illustrated on Fig.~\ref{kpi_sum}: the current status
is shown on the left plot; its approximation to Belle II data is shown
on the right plot.  A violation of the sum rule would be an
unambiguous evidence of a new physics.  With all of ${\cal B}$ and
${\cal A}_{CP}$ results except for ${\cal A}_{CP}(B^0\to K^0\pi^0)$,
the sum rule predicts ${\cal A}_{CP}(B^0\to K^0\pi^0)$ to be
$-0.153\pm0.045$, which is consistent with current measurement.  The
discrepancy can be revealed with higher precision measurements on the
larger statistics of Belle II.  A detailed physics program of the Belle II 
experiment is described in Ref.~\cite{belle2_physics}.

\section{SuperKEKB accelerator and Belle II detector}

The SuperKEKB collider is designed by upgrading the existing KEKB
machine. SuperKEKB should achieve a peak luminosity about
$8\times10^{35}{\rm~cm}^{-2}s^{-1}$. This
will allow to accumulate 
$50{\rm~ab}^{-1}$ around 2021-2022. This integrated luminosity
corresponds approximately to 50 billion $B\bar B$ pair events.

The formula for the luminosity can be expressed as:
\[
L=\frac{\gamma_\pm}{2er_e}\big(1+\frac{\sigma_y^*}{\sigma_x^*}\big)
\big(\frac{I_\pm\xi_{\pm y}}{\beta_y^*}\big)
\big(\frac{R_L}{R_y}\big).
\]
To achieve the designed luminosity goal the "nano-beam" configuration
has been chosen.  To increase the luminosity by the factor of 40, the
accelerator parameters in the middle term of this expression will be
changed: the beam current $I_\pm$ to be increased by the factor of 2,
and the beam size $\beta_y^*$ to be reduced by the factor of 20.

The Belle spectrometer will have to be upgraded to Belle II detector to
accommodate much higher luminosity and to work efficiently in the
conditions of much higher background level.  The physics goals require
also the improvements in the accuracies in all sub-detector systems.
To improve the vertex resolution the silicon vertex detector will be
replaced with a 2-layer DEPFET pixel detector and a 4-layer silicon
strip detector.  The Belle drift chamber will be replaced with a new
one with smaller cell size to cope with the higher occupancy.
Particle identification will be provided by a Time-of-Propagation
(TOP) counter in the barrel region and a proximity focusing Cherencov
ring imaging counter with aerogel radiators in the forward endcap
(ARICH).  Electromagnetic calorimeter will be equipped with a new
electronics with wave-form sampling.  The first two layers (closest to
the interaction region) of the barrel part and the entire endcaps of
the Belle muon system in the flux return of the magnet based on
resistive plate chambers will be replaced with scintillator strips.

\section{Summary}

After eleven years of the successful work the Belle experiment was
stopped.  The Belle detector was partially disassembled and prepared
for its upgrade to Belle II.  The Belle II detector is designed, all
sub-systems will be upgraded or replaced with new ones with better
performance and stability against higher background.  The aim of the
new facility is to achieve a $\Upsilon(4S)$ data set equivalent to
$50{\rm~ab}^{-1}$ (about 50 billion $B\bar B$ pair events) around the
year 2022.  In the accelerator machine upgrade scheme, the increase of
the luminosity will be achieved by drastically squeezing the beam size
at the interaction region.  A rich physics program is aimed to search
for a new physics in quark flavor sector.

\end{document}